\documentclass[aip,twocolumn,pra,showpacs]{revtex4}

\usepackage{epsfig}
\usepackage{graphics}
\usepackage{isolatin1}
\usepackage{amssymb,amsmath}
\newcommand{\ra}{\rangle}
\newcommand{\la}{\langle}

\newcommand{\HC}{H_{\mathrm{c}}}             
\newcommand{\eins}{1\!\! 1}                  
\newcommand{\ket}[1]{|#1\rangle}             
\newcommand{\bra}[1]{\langle #1|}            
\renewcommand{\v}[1]{\mathbf{#1}}            
\renewcommand{\Im}{\mathrm{Im}}              
\newcommand{\mt}[1]{\mathrm{#1}}             
\newcommand{\beq}{\begin{equation}}
\newcommand{\eeq}{\end{equation}}
\newcommand{\beqa}{\begin{eqnarray}}
\newcommand{\eeqa}{\end{eqnarray}}

\begin{document}

\title{Operator normalized quantum arrival times in the presence of
interactions}

\author{G.~C.~Hegerfeldt}
\author{D.~Seidel}
\affiliation{Institut für Theoretische Physik, Universität Göttingen,
  Tammannstr.~1, 37077 Göttingen, Germany}
\author{J.~G.~Muga}
\author{B.~Navarro}
\affiliation{Departamento de Qu\'{\i}mica-F\'{\i}sica, Universidad del
Pa\'{\i}s Vasco, Apartado Postal 644, 48080 Bilbao, Spain}
   
\begin{abstract}
We model ideal arrival-time measurements for free quantum particles and
for particles subject to an external interaction by means of a narrow and
weak absorbing potential. This approach is related to the operational
approach of measuring the first photon emitted from a two-level atom
illuminated by a laser. By operator-normalizing the resulting
time-of-arrival distribution, a distribution is obtained which for freely
moving particles not only recovers the axiomatically derived distribution
of Kijowski for states with purely positive momenta but is also
applicable to general momentum components. For particles interacting with
a square barrier the mean arrival time and corresponding ``tunneling
time'' obtained at the transmission side of the barrier becomes
independent of the barrier width (Hartman effect) for arbitrarily wide
barriers, i.e., without the transition to the ultra-opaque, classical-like
regime dominated by wave packet components above the barrier.
\end{abstract}
\maketitle

\section{Introduction}
A major open question in quantum theory is how to include time
observables and time measurements in a satisfactory way into the
formalism. Pauli pointed out, in a famous 
footnote \cite{Pauli}, that it is not possible to define 
self-adjoint time operators because of the 
semi-boundedness of the canonically conjugate Hamiltonian $H$
(see \cite{Muga-book,Galapon02} for recent discussions of 
Pauli's argument, its domain of applicability and implications). 
A great deal of effort and ingenuity has been devoted to 
pinpoint and overcome this and other difficulties,  
in particular for describing 
the {\it time of arrival} of a quantum particle
\cite{Allcock69,Kijowski74,Werner86,Yamada91,Mielnik94,MBM95,BJ96,GRT96,
Giannitrapani97,
DM97,Leon97,Leavens98,MSP98,MLP98,AOPRU98,Halliwell99,
Finkelstein99,Toller99,MPL99,EM99,KW99,Kijowski99,
EM00,Leon-PRA-2000,BSPME00,BEMS00,BEM01a,Baute01,Galapon02,
Wlodarz02,BEM02,Leavens02,
DEHM02,EMNR03,DEHM03,HSM03,Home03}.
For an extensive review up to 2000 see \cite{ML00}. 

In a physically motivated, axiomatic approach, Kijowski
\cite{Kijowski74}  derived an arrival-time distribution for free
particles coming in from the left (or right), i.e. with only one sign
of the momentum. 
For a one dimensional wave packet coming in from the left, with
momentum (wave number) representation  
$\tilde{\psi}(k)$, $k>0$, at time $t=0$, and the arrival point at 
$x=0$, Kijowski's distribution is given by
\beq \label{PiK}
\Pi_K(t)=\frac{\hbar}{2\pi m}\left|
\int_0^\infty dk\,\tilde{\psi}(k)\sqrt{k}\,e^{-i\hbar k^2t/2m}\right|^2~.
\eeq
Much more recently, this 
distribution has been related 
\cite{Giannitrapani97,MLP98, EM99, ML00} to the positive
operator-valued measure (POVM) generated by the eigenstates of the 
Aharonov-Bohm (maximally symmetric) time-of-arrival operator
\cite{Aharonov-PR-1961},
and generalized for 
systems with interaction \cite{BSPME00,Baute01} and multi-particle states
\cite{BEM02}.
As pointed out in several reviews 
\cite{ML00,Muga-book,EMNR03}, 
one of the important pending questions of Kijowski's
distribution and its generalizations 
is  
the absence of an operational interpretation, in terms of 
some measuring procedure. 
This problem becomes particularly acute when interpreting the puzzling
results obtained for complicated cases 
such as ({\it i}) freely moving particles with wave components coming from 
both sides to the arrival position,
and ({\it ii}) particles subject to some interaction potential
\cite{Leavens02,EMNR03}.   

Another research thread on quantum arrival times is based on  
modeling the irreversible detection by means of effective absorbing
complex potentials. 
They were first used by Allcock in this context \cite{Allcock69}.  
He chose an imaginary potential step and   
found that a strong potential (therefore a 
short detection time) lead to important reflection, i.e.,
many particles 
of the quantum ensemble associated with the incident wave packet
could not be detected; this affected more intensely the slow 
end of the momentum distribution 
and 
distorted the measured signal. In the opposite limit,
i.e., for a very weak absorbing potential, the reflection was avoided but at the price of 
inducing a very large detection delay. 
By deconvolving the 
absorption-time probability density with the 
absorption-time distribution for a particle immersed at rest
in the complex potential, 
in the limit of vanishing absorption, he could define an ``ideal distribution''
which 
turned out to be equal to the flux and therefore not positive definite,
even for states formed only with positive momentum components.   
Kijowski's distribution for positive-momentum states 
emerged 
as the best positively defined fit to this ``ideal distribution''
\cite{Allcock69}.  

Twenty five years later, further investigations of the quantum arrival time 
using complex potentials revealed that 
perfect and fast detection were compatible 
by means of properly designed 
complex potential 
profiles \cite{MBM95,MPL99}. However, an ambiguity in the 
detailed form of these potentials remained, so it was not clear 
how to define, within the complex potential approach, 
an ideal time-of-arrival distribution in an arbitrary 
case. The distributions obtained with 
potentials constructed to absorb perfectly in the  
energy range of a given incident wave packet    
were in general close but not equal to Kijowski's distribution, 
and the relation with actual measurement procedures
remained vague. 
A model proposed by Halliwell of time-of-arrival 
detection by means of an unstable two-level detector
made such a relation more plausible \cite{Halliwell99} (he  
derived a complex potential, again with a step 
form, from the original full set of equations for the two-level
system), but the model lacked a precise physical
content since  
the nature of the actual system or coupling mechanism
were not provided.   
  
An important step forward to relate ``ideal'' and ``operational''
distributions has been the 
development of a  
model, using the ``quantum jump'' technique \cite{qj},
for the detection of first photons emitted by atoms 
illuminated by a laser beam localized in a semi-infinite
\cite{DEHM02,NEMH03} or 
finite region of space \cite{DEHM03}. 
Closely resembling Allcock's results, it was found that 
strong driving by the laser 
caused reflection, whereas  the weak driving regime could be used   
to get the flux by deconvolution from the first-photon distribution, 
but Kijowski's distribution was not obtained in any limit \cite{DEHM02}.
The analogy with Allcock´s work was not accidental since in the  
limit where the inverse life-time $\gamma$ and the Rabi frequency $\Omega$ 
become very large, but keeping the ratio $\Omega^2/\gamma$ 
constant, and $\gamma/\Omega \gg 1$,   
it is possible to find a closed  
equation for the atomic ground-state 
amplitude in terms of an effective complex potential \cite{NEMH03,HSM03}.
For the semi-infinite
laser model, which is physically valid for slow enough atoms \cite{DEHM03},
and the laser 
tuned at resonance with the atomic transition,
it takes the 
simple form $-i\hbar\Omega\Theta(x)^2/(2\gamma)$, where $\Theta(x)$ is the 
Heaviside function, and the laser illuminates the region  $x\ge 0$.
In other words, 
once again, a purely imaginary absorbing  step potential arises, but now with 
a clearcut physical content in terms of laser and atomic parameters.       

As pointed out before, though, the atom-laser model alone did not lead to 
Kijowski's distribution.
In a very recent article,  
Kijowski's distribution has been obtained for freely moving states
with positive momentum components \cite{HSM03}, by applying 
the ``operator normalization'' proposed by Brunetti and Fredenhagen 
\cite{BF02} to the model that describes
the interaction between the atom and a semi-infinite laser 
in the intuitive limit of strong laser field and fast decay
(large $\Omega$ and $\gamma$ respectively).
   
From the previous discussion it should be clear that the problem of
undetected atoms,  
i.e., atoms reflected (or transmitted in the finite beam case) without
emitting  any photon has to be faced in some way. For example in \cite{DEHM02}
``normalization by hand'' for the fraction of detected atoms 
was used for strong driving conditions. Instead,     
for states with incident positive momenta the strategy of the
``operator normalization'' amounts to enhance the 
amplitude of each incident momentum component 
by an amount inversely proportional to the square root of the 
corresponding detection 
probability.
%
%
Operator normalization may be applied whether
or not the laser-atom parameters allow for a 
simplified complex-potential description, 
but the advantadge of such a parameter domain is that a much 
simpler one-channel calculation may be carried out; 
this is particularly useful to tackle 
the more complicated physical situations
examined in the present work.   

One obvious limitation of the semi-infinite models (with the laser 
in explicit form or  
with a complex potential)
is that it is 
not possible to study with them the arrival at a point, say $x=0$, 
corresponding to a state incident from both sides, $x>0$ and $x>0$, 
with negative and positive momenta respectively; similarly, if one is 
interested in the arrivals in the midst of a potential interaction
region, the semi-infinitely extended measurement will severely
affect the dynamics
of the unperturbed system on one side.      
The aim of this paper is to apply, instead of the semi-infinite interaction,
a weak and narrow, minimally perturbing  absorbing complex potential
combined with ``operator normalization'' to these two more elusive
arrival-time problems.  

The motivation for looking at the first case (free motion) 
is a question originally 
posed by Leavens \cite{Leavens98,ML00}: When positive and negative
momenta are present in the wavefunction, Kijowski's original axiomatic
approach does not apply and he obtained, in a heuristic way, a
distribution which in one dimension is given by 
a sum of independent contributions. 
The immediate consequence of this structure is that symmetrical and
antisymmetrical states with respect to the arrival point $x=0$,
formed by adding or substracting two waves with opposite momenta, 
but identical otherwise, 
have {\it the same} arrival distribution, i.e., the distribution is 
``blind'' to the interference between the positive and negative momentum
components and predicts arrivals for the antisymetric case  even where the
wave function vanishes at all times. This result may be 
understood formally from the resolution of the identity in terms of 
eigenstates of the time-of-arrival operator, 
and corresponds to 
a projection of the state onto the positive or negative 
momentum subspaces \cite{ML00},
but a hint on how that process could be implemented 
was missing \cite{EMNR03}. We shall obtain this
distribution by using the weak and narrow absorbing potential, i.e.,
an appropriate 
laser interaction, combined with ``operator normalization''. 
  
The analysis of arrival times when the particle interacts with a potential 
has been
problematic, too. It took a long time to  
generalize Kijowski's distribution in this case 
because the original axiomatic method  
is not applicable 
\cite{DM97, Leon-pre-99,  BSPME00, Leon-PRA-2000,
Baute01}.  
The proposed generalizations, though, 
are purely formal in character. In principle, a narrow and
weak laser 
may be used to excite and detect the atom with minimal disturbance, 
even when the system's motion is affected by an additional 
interaction potential; by compensating the detection losses
by ``operator normalization'' 
a physically motivated 
generalization is obtained. As before,  for simplicity, a complex
potential will play  the role of the laser.  
In this paper we apply the method to tunneling across
the paradigmatic square 
barrier potential, and obtain an arrival time distribution at the 
transmission side. 
It 
is similar to Kijowski's expression, but
modified by the phase of the transmission amplitude. The
effect of operator normalization is that 
the mean
arrival time becomes the average of the ``phase times'' with respect to
the initial  
quantum state, instead of the transmitted wave packet.
As a consequence, the Hartman effect \cite{Hartman-JAP-1962},
namely, the essential 
independence of the average arrival time with respect to the barrier width,
is obtained for arbitrarily large barriers, without the transition 
to an ultra-opaque, classical-like  regime 
dominated by over-the-barrier momentum components
\cite{Hartman-JAP-1962,MEDD02}.


\section{Absorbing potentials and quantum arrival time
distribution} 

Let us consider an
asymptotically free, moving wave packet impinging on an imaginary
potential $-iV_{\epsilon}$ which is located in $-\epsilon\leq x \leq
\epsilon$ and a real potential $U$ localized between $a$ and $b$. 
The Hamiltonian is given by
\begin{equation} \label{eq:H}
  H_{\mt{c}} = \frac{\hat{p}^2}{2m} - iV_{\epsilon} \chi_{\epsilon}(\hat{x})
+U \chi_{[a,b]}(\hat{x})
\end{equation}
where $\chi_{[a,b]}$ is one in $[a,b]$ and zero elsewhere, and
\begin{equation}
  \chi_{\epsilon}(x) = \left\{  \begin{array}{cc}
    1 & : -\epsilon\leq x \leq \epsilon\\
    0 & :\,\mt{elsewhere} \end{array} \right. .
\end{equation}
The scaling of the potential $V_{\epsilon}$ in the limit $\epsilon\to
0$ is given by a function $c(\epsilon)$ and 
\begin{equation}
  V_{\epsilon} = \frac{V_0 L_0}{2c(\epsilon)}.
\end{equation}
Here $V_0$ and $L_0$ are
some arbitrary initial values for the potential height and potential
width. Depending on the structure of $c(\epsilon)$, we may
investigate two
cases: 

(a)~$c(\epsilon)=\epsilon$,

(b)~$c(\epsilon)\sim \epsilon^\alpha~,~~~~~0<\alpha<1$ .\newline
The case (a) yields in the limit $\epsilon\to 0$ 
a delta-function potential with ``strength'' 
$V_0L_0$, $V_{\epsilon}\chi_{\epsilon}(\hat{x}) \to V_0L_0 \delta(\hat{x})$,
whereas case (b) implies a weaker and less perturbing measurement.  
We shall eventually prefer this later case  
but include an analysis of the first one for completion and comparison. 
In the following we shall always apply the limit $\epsilon\to 0$ and
denote these two cases by $\epsilon\stackrel{(a)}{\to} 0$, 
$\epsilon\stackrel{(b)}{\to} 0$, respectively. 

To obtain the time development under $H_{\mt{c}}$ of a wave packet
which is asymptotically free, one first solves the stationary equation
\begin{equation} \label{eq:stat_SG}
  H_{\mt{c}}\phi_k = E_k\phi_k
\end{equation}
for scattering states with real energy
\begin{equation}
  E_k=\frac{\hbar^2 k^2}{2m}.
\end{equation}
By decomposing an initial state as a superposition of eigenfunctions
of $H_{\mt{c}}$, 
its  time development is obtained. This is easy for an initial 
free wave packet coming in from $x = -\infty$ 
in the remote past. Indeed,
\begin{equation} \label{eq:psi_naiv}
\psi(x,t)= \int_0^\infty dk \,\widetilde{\psi}(k) \,\phi_k 
(x)\,e^{-i \hbar k^2 t/2m}
\end{equation} 
describes the time development of a state which in the
remote past behaves like a free wave packet with
$\widetilde{\psi}(k),~k>0$,
the momentum amplitude it would have at $t=0$.
$\phi_k$ would correspond in that case to the scattering states 
for left incidence. 
In the absence of a real potential $(U=0)$ we shall also be interested 
in states with both positive and negative momenta. The formal treatment for 
symmetrical and antisymmetrical wave components is analogous, as shown later 
with more detail. 

It is convenient to use 
the interaction picture with
respect to $H_0 = \hat{p} ^2/2m$,
\begin{eqnarray}
  \HC^I &=& e^{i H_0 t/\hbar}(\HC- H_0) e^{-i H_0 t/\hbar}\nonumber\\
U^I_{\rm c}(t,t_0) &=& e^{i H_0 t/\hbar}e^{-i \HC (t- t_0)/\hbar}e^{-i H_0
  t_0/\hbar},\nonumber\\&&  
\end{eqnarray}
where $U^I_{\rm c}$ is the conditional time development  corresponding
to $\HC^I$. Then Eq.~(\ref{eq:psi_naiv}) can be written as 
\begin{eqnarray} \label{eq:psi_streu}
\psi_t =e^{-i H_0t/\hbar}\,U^I_{\rm c}(t,-\infty)\ket{\psi},
\end{eqnarray}
where $\ket{\psi}\equiv\int dk \ket{k}\tilde{\psi}(k)$. 
An arrival time distribution in the region $-\epsilon\leq x \leq
\epsilon$ is given by the absorption rate \cite{HSM03}
\begin{eqnarray} \label{eq:Pi(t)}
  \Pi(t) &=& \frac{2V_{\epsilon}}{\hbar} \int_{-\epsilon}^{\epsilon}
  dx~|\psi(x,t)|^2\\
  &=& \int dk\,dk'\, \overline{\tilde{\psi}(k)}
  \tilde{\psi}(k') e^{i\hbar(k^2-k'^2)t/2m} f_{\epsilon}(k,k')\nonumber\\
\end{eqnarray}
with the kernel function
\begin{equation} \label{eq:fkk_def}
  f_{\epsilon}(k,k') = \frac{2V_{\epsilon}}{\hbar}
\int_{-\epsilon}^{\epsilon} dx~ \overline{\phi_k(x)}
  \phi_{k'}(x).
\end{equation}
Now $\Pi(t)$ can be written as an expectation value on incoming states in
the form
\begin{equation}
  \Pi(t) = \bra{\psi}\hat{\Pi}_t\ket{\psi}
\end{equation}
and 
\begin{equation}\label{Pihat}
  \hat{\Pi}_t = \frac{2V_{\epsilon}}{\hbar} U_{\mt{c}}^I(t,-\infty)^{\dagger}
  \chi_{\epsilon}(\hat{x}) U_{\mt{c}}^I(t,-\infty).
\end{equation}
To normalize on the level of operators we define \cite{HSM03,SdaggS}
\begin{equation}
  \hat{B} = \int_{-\infty}^{\infty} d t~ \hat{\Pi}_t = \eins -
  U_{\mt{c}}^I(\infty,-\infty)^{\dagger}U_{\mt{c}}^I(\infty,-\infty)   
\end{equation}
and
\begin{equation} \label{eq:PiON_t_def}
  \hat{\Pi}^{\text{\tiny ON}}_t = \hat{B}^{- 1/2} \hat{\Pi}_t
  \hat{B}^{- 1/2}. 
\end{equation}
In Ref.~\cite{HSM03} it has been shown that $\hat{B}$
is diagonal in  
$k$ space, $\bra{k}\hat{B}\ket{k'} = b(k,k')\delta(k-k')$ (remember that 
the integrals over $k$ run from $0$ to $\infty$), and this
leads to the normalized distribution
\begin{eqnarray}
\label{piont}
  \Pi^{\text{\tiny ON}}(t) &=& \bra{\psi}\hat{\Pi}^{\text{\tiny ON}}_t
  \ket{\psi}\nonumber\\
  &=& \int dk dk~\overline{\tilde{\psi}(k)}
  \tilde{\psi}(k')~b(k,k)^{-1/2}b(k',k')^{-1/2}\nonumber\\
  &&\times~e^{i\hbar
    (k^2-k'^2)t/2m} f_{\epsilon}(k,k').
\end{eqnarray}
The kernel function $b(k,k)$ can be calculated as in
Ref.~\cite{HSM03} or
by the following simple argument. Because of $\int dt~\Pi^{\text{\tiny
    ON}}(t) = 1$, $b(k,k)^{-1/2}$ has to cancel the factors which
arise from integrating $\Pi(t)$ in Eq.~(\ref{eq:Pi(t)}) over $t$,
namely $\frac{2\pi m}{\hbar k}f_{\epsilon}(k,k)$. Therefore one has in the
symmetric form provided by Eq.~(\ref{eq:PiON_t_def}) 
\begin{equation} \label{eq:kernel_B}
  b(k,k)^{-1/2} = \left(\frac{\hbar k}{2\pi m f_{\epsilon}(k,k)}\right)^{1/2}
\end{equation}
and
\begin{eqnarray} \label{eq:PiON(t)}
  \Pi^{\text{\tiny ON}}(t) &=& \frac{\hbar}{2\pi m} \int dk\,dk'\,
  \overline{\tilde{\psi}(k)} \tilde{\psi}(k')
  e^{i\hbar(k^2-k'^2)t/2m} \sqrt{kk'}\nonumber\\ &&\times~
  \frac{f_{\epsilon}(k,k')}{\sqrt{f_{\epsilon}(k,k)f_{\epsilon}(k',k')}}.
\end{eqnarray}
The normalization leads immediately to Kijowski's kernel $\sqrt{kk'}$
in the arrival time distribution, modified by a model dependent term
\begin{equation} \label{eq:Fkk_def}
  F_{\epsilon}(k,k') =
  \frac{f_{\epsilon}(k,k')}{\sqrt{f_{\epsilon}(k,k)f_{\epsilon}(k',k')}}
\end{equation}
which has to be investigated for specific situations in the limit
$\epsilon \to 0$.

\section{Examples: From the free case to tunneling particles}

\subsection{Arrival time distribution for free particles}

\begin{figure}[htbp]
  \centering
  \epsfxsize=8cm  \epsfbox{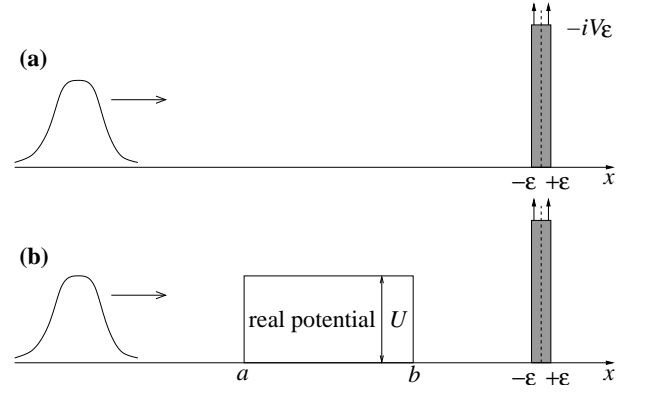}
  \caption{Measurement scheme for the arrival time
    distribution at  $x=0$ of a wave packet coming from the left. (a)
    Free wave packet. (b) Tunneling wave packet. (c) Arrival time
    distribution inside a potential.}
  \label{fig:delta1}
\end{figure}

The easiest case to evaluate Eq.~(\ref{eq:PiON(t)}) is a free incoming
wave packet of positive momenta.
Its Hamiltonian is given by Eq.~(\ref{eq:H}) with $U=0$ and the
general solution of Eq.~(\ref{eq:stat_SG}) in the region $i$,
$i=0,1,2$ associated with $x\leq -\epsilon$, $-\epsilon\leq x \leq
\epsilon$, $\epsilon \leq x$, respectively, is given by
\begin{equation} \label{eq:phi_i}
  \phi_k^{(i)}(x) = \frac{1}{\sqrt{2\pi}} (A_i^+ e^{ik_ix} + A_i^-
  e^{-ik_ix}) 
\end{equation}
where $k_0 = k_2 \equiv k$ and $k_1 \equiv q_{\epsilon} = [k^2 + \frac{im
    V_0 L_0}{\hbar^2 c(\epsilon)}]^{1/2}$ with $\Im\,q_{\epsilon}>0$
(Fig.~\ref{fig:delta1}a).
The amplitudes
$A_i^+$ and $A_i^-$ are determined by the matching conditions at
$x=-\epsilon$ and $x=\epsilon$ and they are explicitly derived in Appendix A. 

{\em (i)  Left incoming states:} The appropriate 
eigenstates have boundary conditions $A_0^+ = 1$ and
$A_2^-=0$. Then one can solve for the other amplitudes 
to 
obtain
\begin{eqnarray}
  A_2^+ &=& e^{-2ik\epsilon}/D \\
  A_0^- &=& -\frac{i}{2}\left(\frac{k}{q_{\epsilon}}
- \frac{q_{\epsilon}}{k}\right) \sin(2q_{\epsilon}\epsilon)
    e^{-2ik\epsilon}/D\\
  A_1^+&=& \frac{1}{2}\left(1+\frac{k}{q_{\epsilon}}\right)e^{-i(k+q_{\epsilon})\epsilon}/D \\
  A_1^- &=& \frac{1}{2}\left(1-\frac{k}{q_{\epsilon}}\right)e^{-i(k-q_{\epsilon})\epsilon}/D
\end{eqnarray}
with the common denominator
\begin{equation}
  D = \cos(2q_{\epsilon}\epsilon) - \frac{i}{2}\left(\frac{k}{q_{\epsilon}} + \frac{q_{\epsilon}}{k}\right)
  \sin(2q_{\epsilon}\epsilon). 
\end{equation}
In the limit $\epsilon\stackrel{(a)}{\to} 0$ one has
\begin{eqnarray}
 A_0^- &\to& -\frac{m V_0L_0}{\hbar^2 k + mV_0L_0}\\
 A_2^+ &\to& \frac{\hbar^2 k}{\hbar^2 k +mV_0L_0}\\
 A_1^+, A_1^- &\to& \frac{1}{2}\,\frac{\hbar^2 k}{\hbar^2 k + mV_0L_0}
  \label{eq:limit}. 
\end{eqnarray}
These results may be checked by considering a $\delta$-potential in 
Eq.~(\ref{eq:H}) from the start.

The limit $\epsilon\stackrel{(b)}{\to} 0$ yields
\begin{eqnarray}
 A_0^- &\to& 0,\\
 A_2^- &\to& 1,\\
 A_1^+, A_1^- &\to& \frac{1}{2}.
\end{eqnarray}
In this limit the measurement region has no effect on the motion of 
the atoms, as seen by Eq.~(\ref{eq:weak_meas}). 
However, by operator normalization 
a finite distribution 
is obtained in both limits. Inserting $\phi_k^{(1)}(x)$ from Eq.~(\ref{eq:phi_i}) into
Eq.~(\ref{eq:fkk_def}) and calculating Eq.~(\ref{eq:Fkk_def}) leads to
\begin{equation}
  F_{\epsilon}(k,k') \to 1,\qquad \epsilon\stackrel{(a)}{\to}
0, \epsilon\stackrel{(b)}{\to}0  
\end{equation}
and the arrival time distribution, $\Pi^{\text{\tiny
    ON}}_{\to}(t)$, for a left incident free
wave packet equals Kijowski's distribution,
\begin{equation}
  \Pi^{\text{\tiny ON}}_{\to}(t) = \Pi_K(t).
\end{equation}

{\em (ii)  The case of general states:} The Hamiltonian $H_{\mt{c}}$ in
Eq.~(\ref{eq:H}) commutes with the parity operator and, as a
consequence, so does the operator $\hat{\Pi}_t$ in
Eq.~(\ref{Pihat}). Therefore its matrix elements between symmetric and
antisymmetric states vanish so that the operator normalization of
$\hat{\Pi}_t$ can be performed  in the subspaces of symmetric and
antisymmetric states independently.

Let us first consider an antisymmetric incident state $|\psi_a\ra$
composed of two identical wave packets with
opposite momenta coming from the right and from the left \cite{Leavens98,ML00}.
We first consider the antisymmetric subspace. The antisymmetric
eigenstates $|k;a\ra$ of $H_0$ and $\phi_k^a$ of $H_{\mt{c}}$ ($k>0$) are
given by
\beq
|k;a\ra\equiv \frac{1}{\sqrt{2}}(|k\ra-|-k\ra)
\eeq
and  an  antisymmetric wavefunction $\psi_a$ can be decomposed as
\beq
|\psi_a\ra=\int_0^\infty dk\,|k;a\ra\la k;a|\psi_a\ra.
\eeq
The antisymmetric eigenstates $\phi_k^a$, of $H_{\mt{c}}$ are obtained with
the boundary conditions  $A_0^+ = 1$ and $A_2^- = -1$, which gives 
\begin{eqnarray}
A_2^+ = -A_0^- &=&  \frac{2 + i(\frac{k}{q_{\epsilon}} - \frac{q_{\epsilon}}{k}) \sin(2q_{\epsilon}\epsilon)}{D}\,e^{-2ik\epsilon}  \\
A_1^+ = -A_1^- &=& \frac{e^{-ik\epsilon}}{\frac{q_{\epsilon}}{k}\cos(q_{\epsilon}\epsilon)
  -i\sin(q_{\epsilon}\epsilon)}. \label{eq:T1_both}
\end{eqnarray}
The wave function in the presence of the imaginary potential 
can now be written as
\beq
\psi_a(x,t)=\int_0^\infty dk\,\la k;a|\psi_a\ra \phi_k^a(x)\,
e^{-i\hbar k^2t/2m}.
\eeq
In spite of the fact that the wave function vanishes at $x=0$ the operator
normalization  preserves a finite arrival distribution even when the width of
the measurement region contracts to 0. With 
Eq.~(\ref{eq:phi_i}) and Eq.~(\ref{eq:fkk_def}) one has, using 
Eq.~(\ref{eq:Fkk_def}), 
\begin{equation}
  F_{\epsilon}(k,k') \to 1,\qquad \epsilon\stackrel{(a)}{\to}
0, \epsilon\stackrel{(b)}{\to}0,  
\end{equation}
which yields
\beqa
\Pi^{\text{\tiny ON}}_{a}(t)&=&\frac{\hbar}{2\pi m}\left|\int_0^\infty dk\,\la k; a|\psi_a\ra
\sqrt{k}\,e^{-i\hbar k^2t/2m}\right|^2  
\\
&=&\frac{\hbar}{\pi m}\left|\int_0^\infty dk\,\tilde{\psi}_a(k)
\sqrt{k}\,e^{-i\hbar k^2t/2m}\right|^2.  \label{Pia}
\eeqa
In the last line we have changed to the ordinary momentum representation 
and have taken the antisymmetry into account. 

A similar treatment may be applied to a 
symmetric wavefunction $\psi_s$ by using symmetric eigenfunctions
and the result is again of the form of Eq.~(\ref{Pia}) with $\psi_a$
replaced by $\psi_s$. 

An arbitrary state $\psi(k)$ can be written in terms of its
symmetric and antisymmetric part as $\psi(k)= \psi_s(k)+\psi_a(k)$. 
By parity, $\Pi^{\text{\tiny ON}}_{\psi}(t)$ is the sum of the
corresponding symmetric and antisymmetric contribution since the  cross
terms vanish. By means of a trivial calculation the sum can be written as
\begin{equation}
\Pi^{\text{\tiny ON}}_{\psi}(t)=\frac{\hbar}{2\pi m}\sum_\pm \left|
\int_0^\infty dk\,\tilde{\psi}(\pm k)\sqrt{k}\,e^{-i\hbar k^2t/2m}\right|^2. 
\end{equation}
This is the operator-normalized arrival-time distribution for a
general free wavefunction. The
expression has been  proposed in Refs.~\cite{Kijowski74,ML00} on more
heuristic grounds as a generalization of the distribution in
Eq.~(\ref{PiK}) from left (or right) incoming states to general free states.

\subsection{Arrival time distribution for tunneling particles}

A major advantage of the operational fluorescence model for the
determination of arrival time distributions is that, in contrast to
the approach of Kijowski, it is not restricted to  free
particles. This means that arbitrary potentials with bounded support can
be considered and the arrival time distribution in the presence of these
interactions can be calculated.

For simplicity, we consider here the case of a rectangular potential
barrier. The Hamiltonian is given by Eq.~(\ref{eq:H})
where, as before, the arrival at $x=0$ is measured and the additional
real potential $U$ is located in $a\leq x \leq b$. We here
investigate the case with $a\leq b \leq 0$ (Fig.~\ref{fig:delta1}b).

For solving the
stationary Schr\"odinger equation,  Eq.~(\ref{eq:stat_SG}),
the $x$-axis has
to be divided into five regions $i$, $i=0,\dots,4$, corresponding to 
$x\leq a, a\leq x \leq b, b \leq x \leq -\epsilon, -\epsilon \leq x
\leq \epsilon, \epsilon \leq x$, respectively. The general solution in
region $i$ is given by Eq.~(\ref{eq:phi_i}) with $k_0=k_2=k_4\equiv
k$, $k_1\equiv \varkappa=[k^2-2mU/\hbar^2]^{1/2}$ and  
$k_3\equiv q_{\epsilon}=[k^2 +
  \frac{imV_0L_0}{\hbar^2c(\epsilon)}]^{1/2}$. In Appendix A we present the
derivation of the $A_i^\pm$ using transfer matrices.
 
\begin{figure}[htbp]
  \centering
  \epsfxsize=8cm \epsfbox{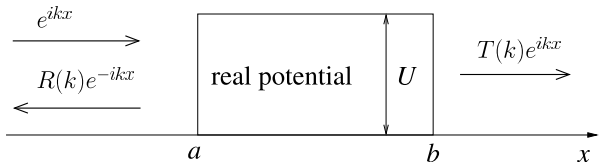}
  \caption{ 
    The stationary scattering function for a wave impinging from the left on 
a real potential barrier. 
$T$ and $R$ are the transmission and reflection amplitudes.} 
  \label{fig:RT}
\end{figure}

In the case of an initial wave packet coming from the left and
crossing the potential region the eigenstates required have 
boundary conditions  $A_0^+ = 1$ and $A_4^- = 0$. Then
one can solve for $A_0^-$ and $A_4^+$ and obtain the amplitudes 
$A_3^\pm$, i.e., the  
solution in the measurement region. In the limit
$\epsilon\stackrel{(a)}{\to}0$ one has with $l=b-a$ and $s=a+b$
\begin{eqnarray}\label{ppp}
  A_3^\pm &\to& \Biggl[ e^{ikl} \left(2 \cos(\varkappa l) -
      i \left(\frac{\varkappa}{k} + \frac{k}{\varkappa}\right) \sin(\varkappa l)
    \right) \nonumber\\
  &&\times~\left( 1+ \frac{m V_0L_0}{\hbar^2 k}\right)\nonumber\\
    && +\, e^{iks}
  \left(\frac{\varkappa}{k} - \frac{k}{\varkappa}\right)\sin(\varkappa
  l) \frac{i m V_0 L_0}{4\hbar^2 k} \Biggr]^{-1}
\end{eqnarray}
whereas in the weak case the limit $\epsilon\stackrel{(b)}{\to}0$ yields
\begin{eqnarray} \label{eq:koeff_limit}
  A_3^\pm &\to& \frac{e^{-ikl}}{2\cos(\varkappa l) -
    i(\frac{\varkappa}{k} + \frac{k}{\varkappa})
    \sin(\varkappa l)}, 
\end{eqnarray}
which may also be obtained
from Eq.~(\ref{ppp}) in the limit $V_0\to 0$. 
The last expression is independent of $V_0 L_0$ and of the  
position of
the potential. It is half the transmission amplitude $T(k)$ of a
rectangular potential barrier without any arrival time measurement, see 
Fig.~\ref{fig:RT}.
Inserting $A_3^\pm$ into $\phi_k^{(3)}$
given by Eq.~(\ref{eq:phi_i}) and
calculating $F_{\epsilon}(k,k')$ by Eq.~(\ref{eq:fkk_def}) and
Eq.~(\ref{eq:Fkk_def}) we obtain
\begin{equation} \label{eq:Fkk_tunnel}
  F_{\epsilon}(k,k') = \frac{\overline{T(k)}
    T(k')}{|T(k)||T(k')|}, \qquad\qquad \epsilon
  \stackrel{(b)}{\longrightarrow} 0,
\end{equation}
and the operator-normalized arrival time distribution of a tunneled
particle takes, with Eq.~(\ref{eq:PiON(t)}), the form
\begin{eqnarray} \label{eq:PiON_tunnel}
  \Pi^{\text{\tiny ON}}_{\mt{pot}}(t) 
  &=& \frac{\hbar}{2\pi m} \Biggl| \int dk~\tilde{\psi}(k) e^{-i\hbar
    k^2 t/2m} \sqrt{k}\,\frac{T(k)}{|T(k)|}\Biggl|^2.\nonumber\\&&   
\end{eqnarray}
The effect of the additional potential is the introduction of  
a phase factor, 
which is the phase of the
transmission amplitude for the real potential in the
weak limit $\epsilon\stackrel{(b)}{\to}0$. 

\begin{figure}[htbp]
  \begin{center}
    \epsfxsize=8cm  \epsfbox{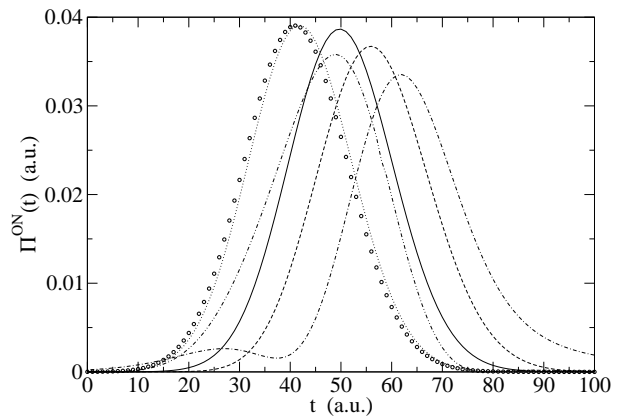}
    \caption{ 
      Arrival time distribution
      $\Pi^{\mt{ON}}$ at $x=0$ in the presence of a potential wall with height
      $U=0$ (solid), $U=0.3$ (dashed), $U=0.48$ (dot-dashed),
      $U=0.58$ (dot-dot-dashed), $U=1.0$ (dotted), $U=2.0$ (circles), 
      potential width $l = 10$,
      for an initial Gaussian wave packet ($x_0 = -50$, 
      $\Delta x = 10, v_0 = 1$) in atomic units $m=\hbar=1$. For
      $U\to\infty$ the mean arrival time approaches the Hartman time.}
    \label{fig:1}
  \end{center}
\end{figure}

The dependence of the arrival time distribution $\Pi^{\text{\tiny
    ON}}_{\mt{pot}}(t)$ on the potential
height $U$ with potential width $l=10$ is shown in
Fig.~\ref{fig:1}. With decreasing $U$ the arrival time at $x=0$ of a particle
starting at $\langle x\rangle = x_0 < 0$ with mean velocity $\langle v
\rangle = v_0$ is 
first delayed but, for increasing potential strength,  
an asymptotic distribution is reached with a mean arrival time
(``Hartman time'') $t_H =
(|x_0|-l)\langle v^{-1}\rangle$ which is smaller than the free arrival
time $|x_0|\langle v^{-1}\rangle$. This is related to the Hartman effect
\cite{Hartman-JAP-1962}, which is discussed in detail in
Section IV. The effect can indeed be observed
\cite{Steinberg-PRL-1993} and it is
complete in the limit $U\to\infty$, corresponding
to an arrival point located in a ``forbidden'' region, in which case
one has
\begin{equation}
  F_{\epsilon}(k,k') \to e^{i(k-k')l},\qquad\qquad U\to \infty.
\end{equation}
In the above setup an incoming free particle was prepared far to
the left and then interacted with an  external real potential. A different
situation arises when one includes the real potential as part of the
preparation procedure through which the particle passes far away on
the left and then continues to propagate freely. 
In this case the incident state 
for operator normalization purposes would be the
normalized transmitted wave packet. 
Formally, this packet may be formed by a projection onto 
a large region to the right of the potential. 
For the positive
results, i.e. transmissions, the normalized incoming free state is
then characterized by  
$T(k)\tilde{\psi}(k)/(\int dk~|T(k)\tilde{\psi}(k)|^2)^{1/2}$ instead of
$\tilde{\psi}(k)$. 
%
%
Applying Kijowski's distribution to the incoming
free state thus prepared gives 
\begin{eqnarray} \label{eq:PiON_Leon}
  \Pi_K^N(t) &=& \frac{\hbar}{2\pi m}~\Big| \int dk\,\tilde{\psi}(k)
  e^{-i\hbar k^2 t/2m}\sqrt{k}\,\,T(k) 
\Big|^2\nonumber\\&&\times~ \left(\int 
dk~|T(k)\tilde{\psi}(k)|^2\right)^{-1}.
\end{eqnarray}
This expression coincides with the proposal of Refs.
\cite{Leon-PRA-2000,DM97,Baute01} for a generalized Kijowski distribution in 
the presence of an external potential and it is thus seen to be related
to our result by a state preparation procedure that selects the transmitted 
particles. 

{\em Remark:} The transmission probability through a potential barrier
as in Fig.~\ref{fig:RT} is given by $\int dk\,|\tilde{\psi}(k)T(k)|^2$ and
one may argue that in this case the total arrival-time probability
should equal  this transmission probability. Instead of
$\Pi^{\text{\tiny ON}}_{\mt{pot}}(t)$ one would then have a modified
distribution,  $\Pi^{\widetilde{\text{\tiny
      ON}}}_{\mt{pot}}(t)$, satisfying
\begin{equation} \label{eq:T_norm}
  \int_{-\infty}^{\infty} dt\,\Pi^{\widetilde{\text{\tiny
        ON}}}_{\mt{pot}}(t) = 
  \int dk\,|\tilde{\psi}(k)T(k)|^2.
\end{equation}
In terms of operators this would require an operator,
$\hat{\Pi}^{\widetilde{\text{\tiny ON}}}_t$, satisfying
\begin{equation}
  \int_{-\infty}^{\infty} dt\,\hat{\Pi}^{\widetilde{\text{\tiny ON}}}_t =
  \int dk\,|T(k)|^2 \ket{k}\bra{k}~.
\end{equation}
With Eqs.~(\ref{eq:kernel_B}), (\ref{eq:Fkk_def}) and
(\ref{eq:Fkk_tunnel})  
it is easily seen
that in this case the kernel of $\hat{B}^{-1/2}$ takes the form
\begin{equation}
  b(k,k)^{-1/2} = \sqrt{\frac{\hbar k}{2\pi m}},
\end{equation}
which yields
\begin{equation}
  F_{\epsilon}(k,k') = f_{\epsilon}(k,k') \to
  \overline{T(k)}T(k'),\quad \epsilon\stackrel{(b)}{\longrightarrow} 0.
\end{equation}
The modified  distribution $\Pi^{\widetilde{\text{\tiny
      ON}}}_{\mt{pot}}(t)$ is then given by
\begin{equation}
 \Pi^{\widetilde{\text{\tiny ON}}}_{\mt{pot}}(t) = \frac{\hbar}{2\pi
   m}~\Big| 
 \int dk\,\tilde{\psi}(k) e^{-i\hbar k^2 t/2m}\sqrt{k}\,\,T(k) \Big|^2,
\end{equation}
which satisfies Eq.~(\ref{eq:T_norm}) and gives the
joint probability density for both  arrival and
transmission. Normalizing this to 1 by hand just yields the distribution
$\Pi_K^N(t)$ of  Eq.~(\ref{eq:PiON_Leon}). Therefore $\Pi_K^N(t)$ can
be understood as 
a conditional probability density for the arrival of the particle 
under the condition that it has been transmitted through the
potential barrier.

\section{Mean arrival times}

To compare our result of tunneling times with previous works, it
is useful to consider not only the arrival time distribution but also 
mean
arrival times. We restrict our analysis to a wave packet
that comes from the far left and collides with a rectangular
potential barrier at $a\leq x\leq b$. The arrival time
behind the barrier is measured at $x=0$ (Fig.~\ref{fig:delta1}b).

With $T(k)=|T(k)|\exp(i\Phi_T(k))$ the mean arrival time, $\langle
t \rangle = \int dt\,t\,\Pi^{\text{\tiny 
    ON}}(t)$, of the distribution in Eq.~(\ref{eq:PiON_tunnel}) is
given by
\begin{equation} \label{eq:mean_t}
 \langle t \rangle = \frac{m}{\hbar}\int
dk~|\tilde{\psi}(k)|^2\frac{|x_0| + \Phi'_T(k)}{k}, 
\end{equation}
where $x_0<0$ denotes the initial value for the mean position
of the wave packet.

This result for $\langle t \rangle$ can be understood as the average
of the ``phase times'' (the time required for a freely moving particle 
plus Wigner's time delay 
\cite{Wigner-PR-1955}) over the initial 
state. In contrast, previous
proposals for the mean arrival time are written in terms of an average
over the {\it transmitted} state \cite{Hartman-JAP-1962,
BSM94, Leon-PRA-2000, MEDD02}, which is just the
first moment of 
the arrival time distribution of Eq.~(\ref{eq:PiON_Leon}). 
These results are not contradictory but
correspond to different state preparations. 

\begin{figure}[htbp]
  \centering
  \epsfxsize=8cm \epsfbox{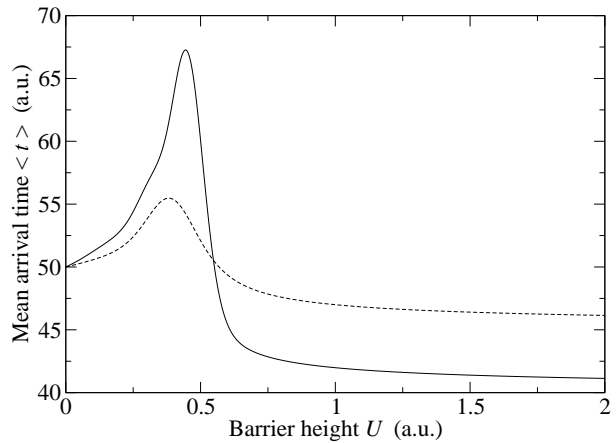}
  \caption{ 
    The mean arrival time $\langle t \rangle$ of Eq.~(\ref{eq:mean_t})
    behind a barrier
    of fixed width $l=10$ (solid), $l=5$
    (dashed) as a function of the
    barrier height $U$. The initial wave packet is a minimal
    uncertainty Gaussian with
    $x_0 = -50$, $v_0 = 1$, $\Delta x = 10$ in atomic units $m=\hbar=1$.
    For $U\to 0$ one
    has the free arrival time $|x_0|\la v^{-1}\ra$, and for $U\to\infty$ the
    arrival time approaches the Hartman time $t_H = (|x_0| -
    l)\la v^{-1}\ra$.} 
  \label{fig:mean_U}
\end{figure}

The dependence with the potential height
is shown in Fig.~\ref{fig:mean_U}, where we have plotted 
$\langle t \rangle$ versus $U$ for fixed barrier width $l$. 
In the free limit $U\to 0$, $\langle t \rangle$ approaches an ``averaged
free arrival time'', $|x_0|\langle v^{-1} \rangle$, and for $U\to
\infty$ it approaches the ``Hartman time'' $(|x_0|-l)\langle v^{-1}
\rangle$, where $\langle v^{-1} \rangle = \int\,dk\,|\tilde{\psi}(k)|
mk^{-1}/\hbar$.

For analyzing the dependence with the barrier width $l$ 
let us heuristically define a ``mean tunneling time'' as an easily calculable 
quantity obtained by  
subtracting from 
$\langle t\rangle$ the classical
time for crossing the non-potential region with average momentum $k_0$,  
\begin{equation}
  \tau = \langle t \rangle - \frac{m(|x_0| - l)}{\hbar k_0}.
\end{equation}
%
%
%
The Hartman effect occurs when $\tau$ in a
tunneling collision of a quantum particle with an opaque square
barrier becomes essentially independent of the barrier width $l$
\cite{Hartman-JAP-1962}.  

\begin{figure}[htbp]
  \centering
  \epsfxsize=8cm \epsfbox{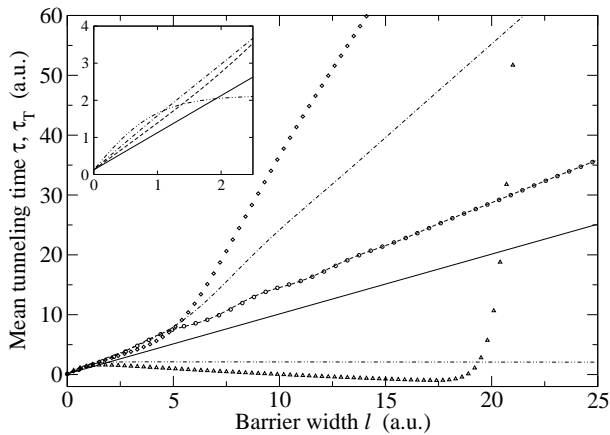}
  \caption{Mean tunneling time $\tau$ (lines) of Eq.~(\ref{eq:mean_t})
    versus $l$ compared with the tunneling time $\tau_T$ (symbols) derived with
    Eq.~(\ref{eq:PiON_Leon}) for different
    barrier heights $U=0$ (solid), $U=0.25$ (dashed, circles), $U=0.48$
    (dot-dashed, diamonds) and $U=1.0$ (dot-dot-dashed,
    triangles). The initial wave packet is a minimal uncertainty Gaussian with
    $x_0 = -50$, $v_0 = 1$, $\Delta x = 10$ in atomic units $m=\hbar=1$.}
  \label{fig:tunnel_l}
\end{figure}

This is shown in Fig.~\ref{fig:tunnel_l}, were we have plotted the
tunneling time $\tau$ versus the potential width $l$. For thin
barriers $\tau$ is above the free traversal time, as shown
in the inset, but for
increasing $l$ there is a sudden transition from a positive delay to a
negative one for increasing $U$. It is clearly visible, that in the
negative delay regime the tunneling time is nearly constant for
increasing $l$ (Hartman regime). In contrast we have plotted the
tunneling time with 
respect to the arrival time distribution $\Pi_K^N(t)$ of
Eq.~(\ref{eq:PiON_Leon}), $\tau_T$, which is related to the proposals of
Ref.~\cite{Leon-PRA-2000,DM97,Baute01}.
Here the behavior of $\tau$ for thin barriers is 
similar, but for increasing $l$ the tunneling time first slowly decreases
and gets actually negative. This has caused many discussions and
warnings that the interpretation of the ``extrapolated phase times'' as an
actual 
tunneling time for the transmitted particle is unjustified
\cite{DBM95, Hauge-RMP-1989}. This question is
not of our concern here though. 
The point we want to stress is that for very thick barriers  the Hartman effect 
vanishes for $\tau_T$, which for widths larger than a
critical barrier length  
grows linearly \cite{BSM94}.
This
is related to the influence of the exponentially decaying 
$|T(k)|$ in Eq.~(\ref{eq:PiON_Leon}),
which causes a domination
of the above-threshold components of the wave
packet \cite{Leon-PRA-2000,BSM94}.
By contrast, $\tau$, obtained with a different preparation procedure does not
show any transition to the classical-like, ultra-opaque regime. 
The intuitive explanation is that operator normalization compensates for
all detection losses due to the $l$-dependence of $|T(k)|$, so that the result
is never dominated by above-the-barrier components.  
%

\section{Conclusion}


We have shown that quantum arrival time distributions can be defined
in a physically motivated way by means of the absorption of a weak and
narrow absorbing potential and operator normalization. As in
Ref.~\cite{HSM03} the use of the absorbing potential corresponds to
the limit of an operational model, where a two-level atom impinges on
a laser region and emits photons. By compensating reflection and
transmission losses in the absorption probability by normalizing the
corresponding operator function the arrival time distribution of Kijowski,
$\Pi_K(t)$, is recovered for freely moving states with positive
momentum. The same method is applied to free states with arbitrary
momentum components, and the result is a generalization of $\Pi_K(t)$
which agrees with the heuristic arguments of Ref.~\cite{Kijowski74}
and with the proposal of Ref.~\cite{ML00}.

The weak and narrow absorbing potential allows a further
generalization for interacting particles. For arrival in the
transmission region behind a square barrier we have shown that our
approach leads to an arrival time distribution which only depends on
the phase of the transmission coefficient. The mean arrival time is
the average over the initial state of the ``phase time'', so that the
Hartman effect, i.e., independence with respect to the barrier length,
is obtained for arbitrarily wide barriers.

A crucial ingredient for obtaining the present results is the way in
which detection losses  are compensated by  operator normalization. 
Reasonable as this modification may be,
there remains to be seen how this can be interpreted from an
operational point of view, i.e. how such
a major intervention could be  performed in practice. Operator
normalization can be viewed as a  state modification,
$|\psi\ra\to\hat{B}^{-1/2}|\psi\ra$, and we note that the modified
state is not normalized. 
The physical process leading from the original
state to the modified one may be a ``filtering'' preparation
by scattering. However, the unbounded nature of the operator 
$\hat{B}^{-1/2}$ restricts such a method to states with a low-energy
cut-off. The feasibility of the transformation in a more 
general case is an open question and further investigation is required.

\acknowledgments{
We acknowledge Ministerio de Ciencia y Tecnolog\'\i a and DAAD for 
a German-Spanish collaboration Grant.
J.G.~M. and B.~N. acknowledge support by Ministerio de Ciencia y 
Tecnolog\'\i a for the Grants BFM2000-0816-C03-03, 
BFM2003-01003 and Universidad del Pa\'\i s Vasco 
for the Grant 00039.310-13507/2001.}

\begin{appendix}
\section{Transfer matrices}

To calculate the transmission and reflection coefficients of the one
dimensional potential problems of Fig.~1 or similar piecewise constant 
potentials we use the transfer matrix
method of Ref.~\cite{Kalotas-AJP-1991} and define ``matching'' matrices
\begin{equation}
  \v{M}_i(x) = \left(\begin{array}{cc} 
    e^{ik_i x}    & e^{-ik_i x} \\
    k_ie^{ik_i x} & -k_i e^{-ik_i x}\end{array}\right).
\end{equation}
for each section $i$ of constant potential $V_i$, where 
\beq
k_i=[k^2-2mV_i/\hbar^2]^{1/2},
\eeq
with ${\rm{Im}}k_i\ge 0$, and $k_i>0$ for positive $V_i$.   
We shall always use the index $0$ for the leftmost section, and 
number the rest rightwards and consecutively. We shall also 
denote by $x_{i+1}$ the boundary point
between sections $i$ and $i+1$, so that the matching conditions 
take the general form 
\beq
  \v{M}_i(x_{i+1}) {A_i^+ \choose A_i^-} =
\v{M}_{i+1}(x_{i+1}) {A_{i+1}^+
    \choose A_{i+1}^-}.
\eeq
Multiplying both sides by $\v{M}_i^{-1}(x_{i+1})$, 
\beq
{A_{i}^+
\choose A_{i}^-}=\v{T}(i,i+1){A_{i+1}^+ \choose A_{i+1}^-}
\eeq
where 
the transfer matrix connecting the amplitudes of regions 
$i$ and $i+1$ is given by   
\beq
\v{T}(i,i+1)\equiv \v{M}_i^{-1}(x_{i+1}) \v{M}_{i+1}(x_{i+1})
\eeq
One may similarly obtain transfer matrices for non contiguous 
regions $i$ and $j\ge i+2$ by multiplying all the
intermediate one-step transfer matrices,  
\beq
\v{T}(i,j)=\v{T}(i,i+1)\v{T}({i+1,i+2}).....\v{T}({j-1,j}).
\eeq
In the following subsections we discuss the cases needed for the
present paper. Note
however that the transfer matrix method is 
of general applicability and may be adapted to continuous potentials
leading in that case to 
differential equations for the amplitudes.

\subsection{Free particles}

For the free arrival time problem (Fig.~1a) one has three regions with
$k_0=k_2\equiv k$, $k_1=q_\epsilon$, and matching points 
at $x_1=-\epsilon$ and $x_2=\epsilon$. 
From the matching matrices one obtains 
\begin{widetext}
\begin{equation} \label{eq:def_Meps1}
  \v{T}(0,2)=
  \left(\begin{array}{cc} 
      \left[\cos(2q_{\epsilon}\epsilon) - \frac{i}{2}(\frac{k}{q_{\epsilon}} +
        \frac{q_{\epsilon}}{k}) 
        \sin(2q_{\epsilon}\epsilon)\right]e^{2ik\epsilon}  &
      \frac{i}{2}(\frac{k}{q_{\epsilon}} - 
      \frac{q_{\epsilon}}{k})\sin(2q_{\epsilon}\epsilon) \\
      -\frac{i}{2}(\frac{k}{q_{\epsilon}} -
      \frac{q_{\epsilon}}{k})\sin(2q_{\epsilon}\epsilon)  & 
      \left[\cos(2q_{\epsilon}\epsilon) + \frac{i}{2}(\frac{k}{q_{\epsilon}} +
        \frac{q_{\epsilon}}{k}) 
        \sin(2q_{\epsilon}\epsilon)\right]e^{-2ik\epsilon} 
    \end{array}\right)
\end{equation}
\end{widetext}
and
\begin{equation} \label{eq:def_Meps2}
  \v{T}(1,2)= \frac{1}{2} \left(\begin{array}{cc} 
      (1+\frac{k}{q_{\epsilon}})e^{i(k-q_{\epsilon})\epsilon} &
      (1-\frac{k}{q_{\epsilon}})e^{-i(k+q_{\epsilon})\epsilon}\\  
      (1-\frac{k}{q_{\epsilon}})e^{i(k+q_{\epsilon})\epsilon} & 
      (1+\frac{k}{q_{\epsilon}})e^{-i(k-q_{\epsilon})\epsilon}
      \end{array} \right).
\end{equation}
This second matrix is useful to obtain the wave function in the 
absorbing region. 
In the limits ${\epsilon\stackrel{(a)}{\to} 0}$ and
${\epsilon\stackrel{(b)}{\to} 0}$ one has
\begin{eqnarray}
  \lim_{\epsilon\stackrel{(a)}{\to} 0} \v{T}({0,2})&=& 
\left( \begin{array}{cc}
    1+\frac{mV_0L_0}{\hbar^2 k} & \frac{mV_0L_0}{\hbar^2 k} \\
    -\frac{mV_0L_0}{\hbar^2 k} & 1-\frac{mV_0L_0}{\hbar^2
      k}\end{array}
  \right) \\ 
\label{eq:weak_meas} \lim_{\epsilon\stackrel{(b)}{\to} 0}
\v{T}({0,2})&=& 
{1~~0 \choose 0~~1}\label{eq:M_identity}\\
\lim_{\epsilon\stackrel{(a)}{\to} 0} \v{T}({1,2}) &=&
\frac{1}{2}{1~~1 \choose 1~~1}\label{eq:Mweak1}\\
\lim_{\epsilon\stackrel{(b)}{\to} 0} \v{T}({1,2}) &=&
\frac{1}{2}{1~~1 \choose 1~~1} \label{eq:Mweak2}.
\end{eqnarray}
For waves incident from the left ($x<0$) the boundary 
conditions are $A_0^+=1$ and $A_2^-=0$ and
there results $A_2^+ =
[T(0,2)_{11}]^{-1}$, $A_0^-= T(0,2)_{21} A_2^+$, $A_1^+
= T(1,2)_{11} A_2^+$ and $A_1^- = T(1,2)_{21} A_2^+$. 

\subsection{Tunneling particles}

In the case of a tunneling particle (Fig.~1b) with $a<b<0$ one has to
consider boundary conditions at $x_1=a$, $x_2=b$, $x_3=-\epsilon$ and
$x_4=\epsilon$. 
$\v{T}(2,4)$ is the same as the transfer matrix  
obtained in the previous section, Eq.~(\ref{eq:def_Meps1}),    
for the regions on both sides of the 
absorbing potential so we only need to calculate $\v{T}(0,2)$,
\begin{widetext}
\begin{equation}
  \v{T}(0,2)= 
  \left(\begin{array}{cc}  
      e^{ikl}\{ \cos(\varkappa l) - \frac{i}{2}
      (\frac{\varkappa}{k} + \frac{k}{\varkappa})
      \sin(\varkappa l)\} &
      -\frac{i}{2} e^{-iks} (\frac{\varkappa}{k} - \frac{k}{\varkappa})
      \sin(\varkappa l) \\
       \frac{i}{2} e^{iks} (\frac{\varkappa}{k} - \frac{k}{\varkappa})
      \sin(\varkappa l) & 
      e^{-ikl}\{ \cos(\varkappa l) + \frac{i}{2}
      (\frac{\varkappa}{k} + \frac{k}{\varkappa})
      \sin(\varkappa l)\} 
    \end{array}\right) 
\end{equation}
\end{widetext}
where $l = b-a$ and $s = a+b$. Since we are interested in the wave
function inside the measurement region $(-\epsilon,\epsilon)$, the
corresponding amplitudes are given from the amplitudes of the rightmost 
region using 
the matrix in Eq.~(\ref{eq:def_Meps2}) as before.

\end{appendix}



\begin{thebibliography}{99}

\bibitem{Pauli}
W.~Pauli, in: S.~Flugge (Ed.), Encyclopedia of Physics, Vol. 5/1, Springer,
Berlin, 1958, S.~60.

\bibitem{Muga-book}
J.G. Muga, R. Sala and I.L. Egusquiza (eds.), 
{\it Time in Quantum Mechanics} (Springer, Berlin, 2002).

\bibitem{Galapon02} 
E. A. Galapon, Proc. Roy. Soc. {\bf 458}, 451 (2002).

\bibitem{Kijowski74}
J. Kijowski, Rep. Math. Phys. {\bf 6}, 362 (1974). 

\bibitem{Allcock69}
G.R. Allcock, Ann. Phys. (N.Y.) {\bf 53}, 253 (1969); {\bf 53}, 286
(1969); {\bf 53}, 311 (1969). 

\bibitem{Werner86} 
R. Werner, J. Math. Phys. {\bf 27}, 793 (1986).

\bibitem{Yamada91} 
N. Yamada and S. Takagi, Prog. Theor. Phys.  {\bf 85},
985 (1991); {\bf 86}, 599 (1991); {\bf 87}, 77 (1992).

\bibitem{Mielnik94}
{B.} {Mielnik},
{Found. Phys.} {\bf 24},
{1113} (1994). 

\bibitem{MBM95}
J.G. Muga, S. Brouard, D. Mac\'\i as,
Ann. Phys. (NY) {\bf 240}, 351 (1995).

\bibitem{BJ96} 
P. Blanchard and A. Jadczyk, Helv. Phys. Acta {\bf 69}, 613 (1996).

\bibitem{GRT96} 
N. Grot, C. Rovelli, R.S. Tate, Phys. Rev. A {\bf 54}, 4676 (1996)

\bibitem{Giannitrapani97}
R. Giannitrapani, Int. J. Theor. Phys. {\bf 36}, 1575 (1997).

\bibitem{DM97} 
V.~Delgado and J.G.~Muga, Phys. Rev. A. {\bf 56}, 3425 (1997). 

\bibitem{Leon97}
J. Le\'on, J. Phys. A {\bf 30}, 4791 (1997). 

\bibitem{MSP98}
J.G.~Muga, R.~Sala, J.P.~Palao, Superlattices Microstruct. {\bf 23},
833 (1998).

\bibitem{Leavens98} 
C.R.~Leavens, Phys. Rev. A {\bf 58}, 840 (1998).

\bibitem{MLP98} 
J.G. Muga, C.R. Leavens and J.P. Palao,
  Phys. Rev. A {\bf 58}, 4336 (1998).

\bibitem{AOPRU98}
Y.~Aharonov, J.~Oppenheim, S.~Popescu, B.~Reznik, and W.~G. Unruh,
Phys. Rev. A {\bf 57}, 4130 (1998).

\bibitem{Halliwell99} 
J.J. Halliwell, Prog. 
Theor. Phys. {\bf 102} 707 (1999).

\bibitem{Finkelstein99} 
J. Finkelstein, Phys. Rev. A {\bf 59}, 3218 (1999).

\bibitem{Toller99} 
M. Toller, Phys. Rev. A {\bf 59}, 960 (1999).

\bibitem{MPL99} 
J.G. Muga, C.R. Leavens, J.P. Palao, Phys. Lett. A
  {\bf 253},21 (1999). 

\bibitem{EM99} 
I.L. Egusquiza and J.G. Muga, Phys. Rev. A {\bf 61},
012104 (1999); {\bf 61}, 059901 (E) (2000).

\bibitem{KW99} 
P. Kocha\'nski and K. W\'odkiewicz, Phys. Rev. A {\bf 60},
2689 (1999).

\bibitem{Kijowski99} 
J. Kijowski, Phys. Rev. A {\bf 59}, 897 (1999).

\bibitem{EM00} 
I.L. Egusquiza and J.G. Muga, Phys. Rev. A {\bf 62},
032103 (2000). 

\bibitem{Leon-PRA-2000}
J.~Le\'on, J.~Julve, P.~Pitanga, F.J.~de~Urr\'{\i}es, Phys. Rev. A
{\bf 61}, 062101 (2000); see also quant-ph/0008025.

\bibitem{BSPME00}
A.D.~Baute,R.S.~Mayato, J.P.~Palao, J.G.~Muga, I.L.~Egusquiza,
Phys. Rev. A {\bf 61}, 022118 (2000). 

\bibitem{BEMS00}
A.D. Baute, I.L. Egusquiza, J.G. Muga, R. Sala-Mayato, Phys. Rev. A
{\bf 61}, 052111 (2000).

\bibitem{BEM01a}
A.D.~Baute, I.L.~Egusquiza, J.G.~Muga, Phys. Rev. A {\bf 64}, 014101 (2001).

\bibitem{Baute01}
A.D. Baute, I.L. Egusquiza, J.G. Muga, {Phys. Rev. A} {\bf{64}},
012501 (2001).  

\bibitem{Wlodarz02} 
J.J. Wlodarz, Phys. Rev. A {65}, 044103 (2002).

\bibitem{BEM02} 
A.D. Baute, I.L. Egusquiza, J.G. Muga, Phys. Rev. A 
{\bf 65}, 032114 (2002).

\bibitem{Leavens02}
C.R.~Leavens, Phys. Lett. A {\bf{303}}, 154 (2002).

\bibitem{DEHM02}
J.A.~Damborenea, I.L.~Egusquiza, G.C.~Hegerfeldt, J.G.~Muga,
Phys. Rev. A {\bf 66}, 052104 (2002).

\bibitem{EMNR03} 
I.L.~Egusquiza, J.G.~Muga, B.~Navarro and A.~Ruschhaupt,
Phys. Lett. A {\bf 313}, 498 (2003).

\bibitem{DEHM03}
J.A.~Damborenea, I.L.~Egusquiza, G.C.~Hegerfeldt, J.G.~Muga,
J. Phys. B: At. Mol. Opt. Phys. {\bf 36}, 2657 (2003).

\bibitem{HSM03} 
G.C. Hegerfeldt, D. Seidel, J.G. Muga, 
Phys. Rev. A {\bf 68}, 022111 (2003).

\bibitem{Home03} 
A. Manirul, A. S. Majumdar, D. Home, S. Sengupta, 
Phys. Rev. A {\bf 68}, 042105 (2003).

\bibitem{ML00} 
J.G. Muga and C.R. Leavens, Phys. Rep. {\bf 338}, 353 (2000).

\bibitem{Aharonov-PR-1961}
Y. Aharonov, D. Bohm, Phys. Rev. {\bf 122}, 1649 (1961).

\bibitem{qj}  G.C. Hegerfeldt and T.S. Wilser, in:
{\it Classical and Quantum Systems.} 
Proceedings of the Second International Wigner Symposium, July
1991, edited by H.D. Doebner, W. Scherer, and F. Schroeck, (World
Scientific, Singapore, 1992), p. 104;
G.C. Hegerfeldt,
\newblock Phys. Rev. A {\bf 47}, 449 (1993); G.C. Hegerfeldt and
D.G. Sondermann, Quantum 
  Semiclass.~Opt.~{\bf 8}, 121 (1996). For a review cf.  M.B. Plenio
  and P.L. Knight, 
Rev. Mod. Phys. {\bf 70}, 101 (1998). The quantum jump approach is
essentially equivalent to the Monte-Carlo wavefunction approach of 
 J. Dalibard, , Y. Castin and  K. M{\o}lmer,  
    Phys. Rev. Lett., {\bf68}, 580 (1992), and to the quantum trajectories of 
 H. Carmichael, {\em An Open Systems Approach to Quantum 
Optics}, Lecture Notes in Physics m18, (Springer, Berlin,  1993).

\bibitem{NEMH03} 
B. Navarro, I.L. Egusquiza, J.G. Muga, G.C. Hegerfeldt, 
J. Phys. B: At. Mol. Opt. Phys. {\bf 36}, 3899 (2003).

\bibitem{BF02} 
R. Brunetti and K. Fredenhagen, Phys. Rev. A {\bf 66}, 044101 (2002).

\bibitem{Leon-pre-99}
J.~Le\'on, J.~Julve, P.~Pitanga, F.J. de Urr\'{\i}es, e-print
quant-ph/9903060.

\bibitem{Hartman-JAP-1962}
T.E. Hartman, J. Appl. Phys. {\bf 33}, 3427 (1962).

\bibitem{BSM94}
S. Brouard, R. Sala, J.G. Muga, Phas. Rev. A {\bf 49}, 4312 (1994).

\bibitem{MEDD02}
J.G.~Muga, I.L.~Egusquiza, J.A.~Damborenea, F.~Delgado, 
Phys. Rev. A {\bf 66}, 042115 (2002). 

\bibitem{Steinberg-PRL-1993}
A.M.~Steinberg, P.G.~Kwiat, R.Y.~Chiao, Phys. Rev. Lett. {\bf 71}, 708 (1993).

\bibitem{Wigner-PR-1955}
E.P.~Wigner, Phys. Rev. {\bf 98}, 145 (1955).

\bibitem{DBM95}
V.~Delgado, S. Brouard, J.G.~Muga, Solid State Commun. {\bf 94}, 979 (1995).

\bibitem{Hauge-RMP-1989}
E.H.~Hauge, J.A.~St{\o}vneng, Rev. Mod. Phys. {\bf 61}, 917 (1989).

\bibitem{Kalotas-AJP-1991}
T.M.~Kalotas, A.R.~Lee, Am. J. Phys. {\bf 59}, 48 (1991).

\bibitem{SdaggS}
$\hat{B}=1-S^\dagger S$ in the language of scattering theory, 
$S$ being the usual scattering operator connecting incoming and outgoing 
asymptotic states. $\hat{B}$ averaged over the incoming asymptotic state
provides the total detection probability.



\end{thebibliography}
\end{document}